\documentclass[reprint,longbibliography,aps]{revtex4-2}

\usepackage{graphicx} 
\graphicspath{figs}
\usepackage{amsmath}
\usepackage{ amssymb }

\usepackage{color}

\usepackage{float} 
\usepackage{wrapfig} 

\usepackage[makeroom]{cancel} 
\usepackage{comment}

\usepackage{lipsum} 

\usepackage{outlines}

\graphicspath{{./figs/}} 

\newcommand{\beq}{\begin{equation}}
\newcommand{\eeq}{\end{equation}}

\newcommand{\bvec}{\begin{pmatrix}}
\newcommand{\evec}{\end{pmatrix}}
\newcommand{\lp}{\left(}
\newcommand{\rp}{\right)}

\newcommand{\ve}[1]{\mathbf{#1}}

\AtBeginDocument{%
\newwrite\bibnotes
\def\bibnotesext{Notes.bib}
\immediate\openout\bibnotes=\jobname\bibnotesext
\immediate\write\bibnotes{@CONTROL{REVTEX42Control}}
\immediate\write\bibnotes{@CONTROL{%
		aip42Control,author="08",editor="1",pages="1",title="0",year="1"}}
\if@filesw
\immediate\write\@auxout{\string\citation{aip42Control}}%
\fi
}%

\begin{document}



\title{The Critical Role of Isopotential Surfaces for Magnetostatic Ponderomotive Forces}

\author{Ian E. Ochs}
\email{iochs@princeton.edu}
\affiliation{Department of Astrophysical Sciences, Princeton University, Princeton, New Jersey 08540, USA}
\author{Nathaniel J. Fisch}
\affiliation{Department of Astrophysical Sciences, Princeton University, Princeton, New Jersey 08540, USA}

\date{\today}

\begin{abstract}
	
By producing localized wave regions at the ends of an open-field-line magnetic confinement system, ponderomotive walls can be used to differentially confine different species in the plasma.
Furthermore, if the plasma is rotating, this wall can be magnetostatic in the lab frame, resulting in simpler engineering and better power flow.
However, recent work on such magnetostatic walls have shown qualitatively different potentials than those found in the earlier, non-rotating theory.
Here, using a simple slab model of a ponderomotive wall, we resolve this discrepancy.
We show that the form of the ponderomotive potential in the co-moving plasma frame depends on the assumption made about the electrostatic potential in the lab frame.
If the lab-frame potential is unperturbed by the magnetic oscillation, one finds a parallel-polarized wave in the co-moving frame, while if each field line remains equipotential throughout the perturbation region, one finds a perpendicularly-polarized wave. 
This in turn dramatically changes the averaged ponderomotive force experienced by a charged particle along the field line: not only its scaling, but also its direction.

\end{abstract}

\maketitle

\section{Introduction}

It has long been known that transverse waves at the ends of an open-field-line plasma can confine species according to a cyclotron-frequency-dependent ponderomotive potential \cite{Gaponov1958PotentialWells,Pitaevskii1961ElectricForces,motzRadioFrequencyConfinementAcceleration1967,Miller2023RFPlugging}, which can even form a one-way wall, acting as a Maxwell Demon \cite{dodinPonderomotiveBarrierMaxwell2004,Dodin2005}.
This potential has been repeatedly confirmed in experiments \cite{Dimonte1982PonderomotivePseudopotential, Anderegg1995LongIon}, including applications for isotope separation \cite{Hidekuma1974PreferentialRadiofrequency,Weibel1980SeparationIsotopes} and 
differential confinement \cite{Hiroe1975RadiofrequencyPreferential,Watari1978RadiofrequencyPlugging}.
Indeed, similar ponderomotive potentials are ubiquitous across different areas of physics, from the operation of free electron lasers in laser physics \cite{Roberson1989ReviewFree}, to trapping near the Rabi frequency in atomic phyics \cite{Savage1996IntroductionLight,Ishihara2021OpticalManipulation}, to the optical tweezers used in biophysics \cite{Ashkin1986ObservationSinglebeam}, and even to the stable driven inverted pendulum of classical mechanics \cite{Landau1976Mechanics}.
Exploiting these ponderomotive forces is of particular interest given the recent resurgence of mirror confinement schemes in fusion physics \cite{Burdakov2016MultiplemirrorTrap,Fowler2017NewSimpler,Beery2018PlasmaConfinement,White2018CentrifugalParticle,Miller2021RateEquations,Egedal2022FusionBeam,Endrizzi2023PhysicsBasis}, especially given its ability to preferentially expel ash species from the plasma, which is particularly important for aneutronic fusion schemes \cite{Magee2019DirectObservation,Putvinski2019,kolmes2022waveSupported,Ochs2022ImprovingFeasibility,Magee2023FirstMeasurements}.

Forming these ponderomotive end plugs has historically required the generation of cyclotron-frequency waves.
However, the increasing use of $\ve{E} \times \ve{B}$ rotation \cite{Lehnert1971} for stabilization \cite{Cho2005,Carter2009,Ivanov2013GasdynamicTrap} and centrifugal confinement \cite{bekhtenev1980problems,Ellis2001,Ellis2005,Teodorescu2010} in mirror systems opens up an alternative possibility: to use a magnetostatic perturbation in the lab frame, which the moving plasma will see as an oscillating electromagnic wave.
As shown in the theory of resonant diffusion of alpha particles, such a scheme can have significant engineering and power flow advantages \cite{fetterman2010stationary,fetterman2012wave}.

Such a magnetostatic ponderomotive wall scheme was recently proposed, and shown to exhibit significant confining potential \cite{Rubin2023MagnetostaticPonderomotive,Rubin2023GuidingCenter}.
However, the form of the ponderomotive potential found in that study unexpectedly and significantly differed from the gyrofrequency-dependent potential familiar from the earlier, non-rotating literature \cite{Gaponov1958PotentialWells,Pitaevskii1961ElectricForces,motzRadioFrequencyConfinementAcceleration1967,Miller2023RFPlugging,dodinPonderomotiveBarrierMaxwell2004,Dodin2005}.

\begin{figure}
	\centering
	\includegraphics[width=\linewidth]{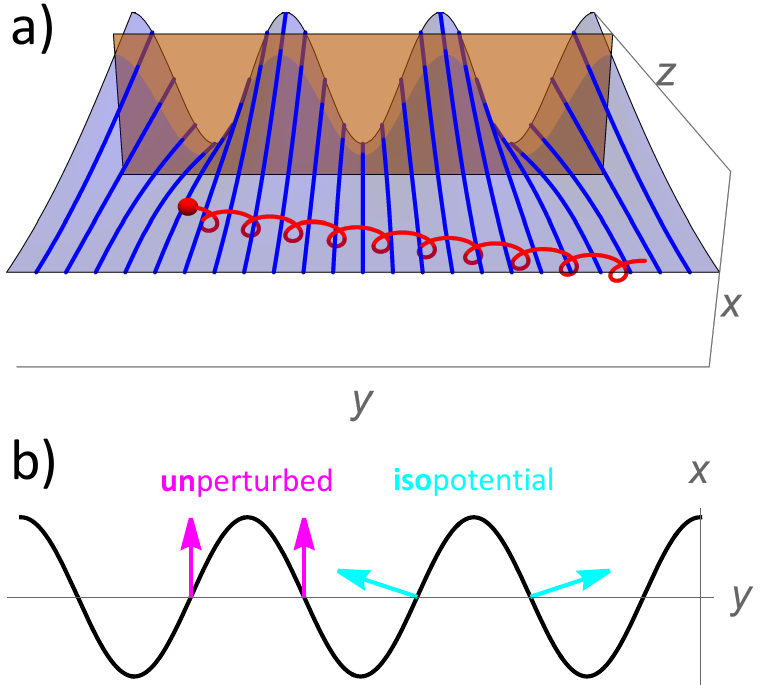}
	\caption{Schematic of the model studied in the text.
	``Radial,'' ``azimuthal,'' and ``axial'' coordinates are represented by $x$, $y$, and $z$, respectively.
	(a) Magnetic field lines (blue lines) along $\hat{z}$  are perturbed by an applied oscillating field along $\hat{x}$, so field lines originally at the same $x$ at $z=0$ will be at different $x$ at $z>0$. 
	At $x = 0$, the electric field points along $\hat{x}$.
	We study the ponderomotive forces along $\hat{z}$ of a charged particle (red sphere) that  both $\ve{E} \times \ve{B}$ drifts along $-\hat{y}$ and streams along $\hat{z}$.
	(b) Projections of two possible models for the electric field at the orange plane in (a). 
	Magenta arrows correspond to the direction of the \textbf{un}perturbed electric field, while cyan arrows correspond to the direction of the field if the perturbed field lines become \textbf{iso}potential surfaces.
	These two models result in very different forces on the charged particle.}
	\label{fig:FieldsSchem}
\end{figure}

In this paper, we identify the root of these discrepancies, by showing that the ponderomotive behavior depends sensitively on the assumptions made about the electric potential structure.
Specifically, we consider a very simple slab analog to the magnetostatic end plug proposed in Ref.~\onlinecite{Rubin2023MagnetostaticPonderomotive}, with a purely radial perturbing magnetic field (Fig.~\ref{fig:FieldsSchem}).
This simple field analytically allows two different assumptions for the behavior of the electric potential during the perturbation: (a) for the potential to remain \textbf{unperturbed} under the influence of the perturbing magnetostatic field, or (b) for the perturbed field lines to become \textbf{isopotential} surfaces.
These two assumptions respectively represent the limit of (a) slow plasma response, where the plasma does not have time to respond to the wave, and (b) fast plasma response, where the electrons move along field lines, shorting the parallel electric field, as underlies magnetohydrodynamic theory and Ferraro's isorotation theorem \cite{Ferraro1937,Northrop1963AdiabaticMotion}.

Using a slab, rather than a cylinder, allows us to apply a very simple Lorentz boost of the lab-frame field, with its associated potential structure, to the plasma rest frame.  
Because the quantity $\ve{E} \cdot \ve{B}$ is Lorentz-invariant, the polarization of the wave in the plasma frame depends on the potential structure in the lab frame: in particular, the wave is parallel-polarized when the original electric potential is unperturbed, and transversely-polarized when the perturbed field lines become the isopotentials.
This difference in wave structure results in dramatic differences in the form of the ponderomotive potential in each case: it changes whether the potential depends on the local value of the perturbed magnetic field or its integral; it changes the scaling behavior with mode number; and it can even change the direction of the resulting force.
Single-particle simulations confirm the theoretical predictions, as well as the limits in which the theory breaks down, including the onset of warm-plasma effects on the ponderomotive potential, as well as the onset of Landau-resonant interactions at high enough wave amplitude \cite{Karney1978,karney1979stochastic}.

\section{Slab Model of Magnetostatic Perturbation}
For simplicity, we consider a slab model of a plasma undergoing $\ve{E} \times \ve{B}$ direction in the $-\hat{y}$ direction, with $\ve{E}_0 = E_0 \hat{x}$ and $\ve{B}_0 = B_0 \hat{z}$ (Fig.~\ref{fig:FieldsSchem}).
To this field, we add a $z$-dependent perturbing magnetic field along the $\hat{x}$ direction, $\ve{\tilde{B}} = \tilde{B}(z) \cos (k y) \hat{x}$.
In the unperturbed case where the magnetic perturbation does not affect the electric field, than we have simply $\ve{E}_\text{un} = \ve{E}_0$.


The form of the perturbing magnetic field we have chosen also makes it easy to solve for the case where the perturbed field lines are isopotential curves, since each field line lies solely in the $x-z$ plane.
At $z = 0$, we assume that the field lines are unperturbed, and we have $\ve{E}(x,y,z=0) = \ve{E}_0$ and $\ve{B}(x,y,z=0) = \ve{B}_0$.
Thus,
\begin{align}
	\phi(x,y,z=0) = - E_0 x. \label{eq:midplanePotential}
\end{align}
Then, for any point, we can trace each field line back to $z = 0$ to determine the potential, by solving:
\begin{align}
	\frac{dx}{dz} = \frac{\ve{B}_x}{\ve{B}_z} = \frac{\tilde{B}(z)}{B_0} \cos (k y).
\end{align}
This gives:
\begin{align}
	x(z=0) = x(z) - \cos (k y) \int_0^z dz' \frac{\tilde{B}(z')}{B_0}. 
\end{align}
Using Eq.~(\ref{eq:midplanePotential}), this gives:
\begin{align}
	\phi_\text{iso} &= - E_0 \left[x(z) - \cos (k y) \int_0^z dz' \frac{\tilde{B}(z')}{B_0}\right],\\
	\ve{E}_\text{iso} &= E_0 \left[\hat{x} + \hat{y}  \int_0^z dz' k \frac{\tilde{B}(z')}{B_0} \sin (ky) - \hat{z} \frac{\tilde{B}(z)}{B_0} \cos (ky) \right].
\end{align}
In contrast to $\ve{E}_\text{un}$, it can be easily verified that $\ve{E}_\text{iso}$ satisfies $\ve{E} \cdot \ve{B} = 0$.
Both fields satisfy $\nabla \times \ve{E} = 0$.

\subsection{Plasma-Frame Fields}
To apply the existing ponderomotive theory \cite{dodinPonderomotiveBarrierMaxwell2004,Dodin2005}, we boost to the co-moving frame, $(x,y,z) \rightarrow (x,y',z)$, where $y' = y + t (c E_0/B_0)$.
Assuming that the plasma flows sub-relativistically ($c E_0/B_0 \ll 1$), the only components that change are the $\hat{x}$ and $\hat{z}$ components of the electric field, according to:
\begin{align}
	\ve{E}' = \ve{E} + \frac{\ve{v}_\text{boost}}{c} \times\ve{B} = \ve{E} + \frac{\left(\ve{E}_0 \times \ve{B}_0 \right) \times \ve{B}}{B_0^2}.
\end{align}

In the ``unperturbed'' case where the magnetic perturbation does not affect the electric field, we have:
\begin{align}
	\ve{E}_\text{un}' &= E_0 \frac{\tilde{B}(z)}{B_0} \cos \left(k y' - \omega t \right) \hat{z}, \label{eq:restFrameEUnperturbed}
\end{align}
where $\omega \equiv k c E_0 / B_0$ is the Doppler-shifted frequency seen by the wave.
Thus, for unperturbed electric field case, the plasma sees a wave in its rest frame with a polarization purely parallel to the background magnetic field, with a strength proporational to the local perturbed magnetic field amplitude.

Meanwhile, when the field lines are isopotentials, the plasma-frame electric field is instead given by:
\begin{align}
	\ve{E}_\text{iso}' &= E_0 \int_0^z dz'' k \frac{\tilde{B}(z'')}{B_0} \sin \left(k y' - \omega t \right) \hat{y}. \label{eq:restFrameIsopotential}
\end{align}
Thus, in this case, the plasma sees a purely perpendicular mode, which is equal in magnitude to the \emph{integrated} amplitude of the perturbing magnetic field.
In other words, the potential structure in the lab frame completely changes the character of the wave in the co-moving frame.

\subsection{Ponderomotive Potentials}
In the plasma frame, the ponderomotive potential (for a cold plasma) is \cite{Dodin2005}:
\begin{align}
	\Phi = \frac{q^2}{4 m \omega^2} \left( \frac{|E_+|^2}{1+\Omega/\omega} + \frac{|E_-|^2}{1-\Omega/\omega} + |E_\parallel|^2\right), \label{eq:coldFluidPonderomotive}
\end{align}
where $q$ and $m$ are the mass and charge of the particle, and $\Omega \equiv q B_0/m c$ is the cyclotron frequency.
Here, $E_\parallel$ is the wave amplitude parallel to $\ve{B}_0$, and $E_+$ and $E_-$ are the amplitudes of the right- and left-handed circularly polarized modes, with polarizations $\boldsymbol{\tau}_{\pm} = \frac{1}{\sqrt{2}}(\hat{x} \pm i\hat{y})$ respectively.
In terms of the linear polarization amplitudes $E_x$ and $E_y$, these circular polarization amplitudes are given by:
\begin{align}
	E_+ &= \frac{1}{\sqrt{2}} \left(E_x - i E_y\right); \quad
	E_- = \frac{1}{\sqrt{2}} \left(E_x + i E_y\right). \label{eq:circularPolarizations}
\end{align}

Thus, the ponderomotive potential for the unperturbed electric field case, using Eq.~(\ref{eq:restFrameEUnperturbed}) in Eq.~(\ref{eq:coldFluidPonderomotive}), is:
\begin{align}
	\Phi_\text{un} &= \frac{q^2 \tilde{B}^2}{4 m c^2 k^2} = \frac{m}{4} \frac{\tilde{\Omega}^2}{k^2}, \label{eq:PhiUnUnnormalized}
\end{align}
where $\tilde{\Omega}$ is the cyclotron frequency associated with the perturbing magnetic field. 
Notice that this reproduces the potential from Ref.~\onlinecite{Rubin2023MagnetostaticPonderomotive}'s Eqs.~(21) and (44).

Meanwhile, the potential for the isopotential field line case, using Eqs.~(\ref{eq:restFrameIsopotential}) and (\ref{eq:circularPolarizations}) in Eq.~(\ref{eq:coldFluidPonderomotive}), is:
\begin{align}
	\Phi_\text{iso} &= \frac{q^2 }{4 m c^2 } \frac{(\int_0^z dz' \tilde{B}(z'))^2}{1-\Omega^2/\omega^2} = \frac{m}{4} \frac{\left(\int_0^z dz' \tilde{\Omega}(z')\right)^2}{1-\Omega^2/\omega^2}. \label{eq:PhiEqUnnormalized}
\end{align}
This potential exhibits several major differences from the unperturbed case.
Instead of scaling with the square of the wavelength $1/k$, the strength of the potential scales with the square of the scale length of the perturbed region $L$.
The potential can also be repulsive or attractive depending on the ratio of the gyrofrequency $\Omega$ to the Doppler-shifted wave frequency $\omega$.
Finally, the potential does not necessarily vanish at the points where the perturbing magnetic field vanishes; it only vanishes if the field line is actively returned to its original position in the $x-y$ plane, by reversing the perturbing field.

\section{Numerical Verification}
To verify the ponderomotive theory, we perform single-particle simulations using Zenitani and Umeda's second-order generalization \cite{Zenitani2018} of the Boris method \cite{boris1970relativistic,Qin2013}, using the LOOPP code developed in Refs.~\onlinecite{ochs2021nonresonant,Ochs2023PonderomotiveRecoil}.
To perform the simulations, we nondimensionalize the equations of motion: 
\begin{align}
	\frac{d\ve{\bar{x}}}{d\bar{t}} &= \bar{v}; \qquad
	\frac{d\ve{\bar{v}}}{d\bar{t}} = \ve{\bar{E}} + \ve{\bar{v}} \times \ve{\bar{B}},
\end{align}
where
\begin{alignat}{5}
	\bar{t} &= \Omega t; & \qquad \ve{\bar{v}} &= \ve{v}/c; & \qquad \bar{\ve{x}} &= \ve{x} \, \Omega/c; \\
	\ve{\bar{E}} &= \ve{E} / B_0; & \ve{\bar{B}} &= \ve{B} / B_0. &&
\end{alignat}
The magnetic field is the same for either assumption about the electric potential structure:
\begin{align}
	\ve{\bar{B}} \; \; &= \hat{z} + \bar{\tilde{B}}(\bar{z}) \cos (\bar{k} \bar{y}) \hat{x},
\end{align}
where $\bar{k} = k c/\Omega$, and $\bar{\tilde{B}} = \tilde{B} / B_0$.
For the unperturbed potential case, the normalized electric field is given by:
\begin{align}
	\ve{\bar{E}}_\text{un} &= \bar{E}_0 \hat{x},
\end{align}
where $\bar{E}_0 = E_0/B_0$. 
For the isopotential field line case, it is given by:
\begin{align}
	\ve{\bar{E}}_\text{iso} &= \bar{E}_0 \left(\hat{x} + \hat{y} \int_0^{\bar{z}} d\bar{z}' \bar{k} \bar{\tilde{B}}(\bar{z}') \sin (\bar{k}\bar{y}) - \hat{z} \bar{\tilde{B}}(\bar{z}) \cos (\bar{k}\bar{y}) \right).
\end{align}

In these dimensionless units, the change in dimensionless parallel kinetic energy of the oscillation center (OC) $\bar{K} \equiv \langle\bar{v}_z \rangle^2/2$ should balance against the dimensionless ponderomotive potential \cite{Dodin2005}, given for the unperturbed and isopotential cases from Eqs.~(\ref{eq:PhiUnUnnormalized}-\ref{eq:PhiEqUnnormalized}) by:
\begin{align}
	\bar{\Phi}_\text{un} &= \frac{1}{4} \frac{\bar{\tilde{B}}^2}{\bar{k}^2};\label{eq:PhiUnNormalized}\\
	\bar{\Phi}_\text{iso} &= \frac{1}{4} \left(\int_0^{\bar{z}} d\bar{z}' \bar{\tilde{B}}(\bar{z}')\right)^2\frac{\bar{\omega}^2}{\bar{\omega}^2-1}.\label{eq:PhiEqNormalized}
\end{align}
Here, $\bar{\omega}$ is the normalized Doppler-shifted frequency, $\bar{\omega} \equiv \omega/\Omega = \bar{k}\bar{E}_0$.
The isopotential ponderomotive potential is attractive for $\bar{\omega} < 1$, and repulsive for $\bar{\omega} > 1$.

As the specific form of the field, we take:
\begin{gather}
	\bar{\tilde{B}} = \bar{\tilde{B}}_0 a \frac{\bar{z}^{a-1}}{(1 + \bar{z}^a)^2}\\
	\int_0^{\bar{z}} d\bar{z}'\bar{\tilde{B}}(\bar{z}) = \bar{\tilde{B}}_0 \bar{L} \frac{\bar{z}^{a}}{1 + \bar{z}^a},
\end{gather}
with $a = 5$.
This field vanishes in the large-$\bar{z}$ limit, but its integral does not, so that the unperturbed ponderomotive potential $\Phi_\text{un}$ goes to 0 as $\bar{z}\rightarrow \infty$, but the isopotential ponderomotive potential $\Phi_\text{iso}$ does not.
For all simulations, we consider a nonrelativistically drifting plasma, $\bar{E} = v_{E\times B} / c = 0.01$, and a particle initialized with $\bar{v}_z = 0.01$ and $\bar{v}_\perp = 0$.
We then  perform four different simulations (labeled A-D), including both the unperturbed and isopotential cases, as detailed in Table~\ref{tab:parameters}.

\begin{table}
	\centering
	\begin{tabular}{c | c | c c c | c}
		Case & Type & $\bar{\tilde{B}}$ & $\bar{k}$ & $\bar{L}$ & $\bar{\omega}$\\
		\hline
		A & Unperturbed & 0.01 & 1 & 300 & 0.01\\
		B & Isopotential & 0.02 & 10 & 5 & 0.1\\
		C & Isopotential & 0.001 & 1030 & 5 & 10.3\\
		D & Isopotential & 0.003 & 530 & 5 & 5.3
	\end{tabular}
	\caption{Parameters that differ for the single-particle simulations in the figures.
	Also shown is the derived quantity $\bar{\omega} = \bar{k} \bar{E}_0$, which determines the sign of $\Phi_\text{iso}$.}
	\label{tab:parameters}
\end{table}

\begin{figure}
	\centering
	\includegraphics[width=\linewidth]{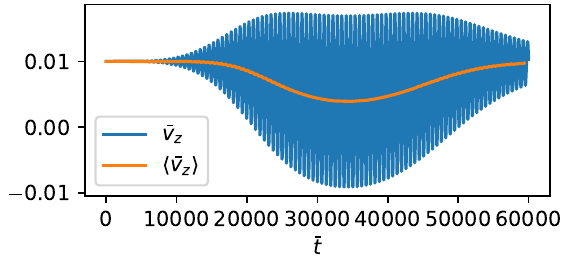}
	\caption{Full parallel velocity $\bar{v}_z$ vs. oscillation center velocity $\langle \bar{v}_z \rangle$ for case A.
	The OC velocity enters the kinetic energy term that balances with the ponderomotive potential $\Phi$.}
	\label{fig:OCSchem}
\end{figure}

To calculate the OC velocity, we time-average over a wavecycle, as shown in Fig.~\ref{fig:OCSchem} for case~A.
This allows us to numerically calculate the normalized OC kinetic energy $\bar{K}$ as a function of $\bar{z}$, and see whether the sum of the numerical kinetic energy and theoretical potential energy from Eqs.~(\ref{eq:PhiUnNormalized}-\ref{eq:PhiEqNormalized}) remains constant.

\begin{figure}
	\centering
	\includegraphics[width=\linewidth]{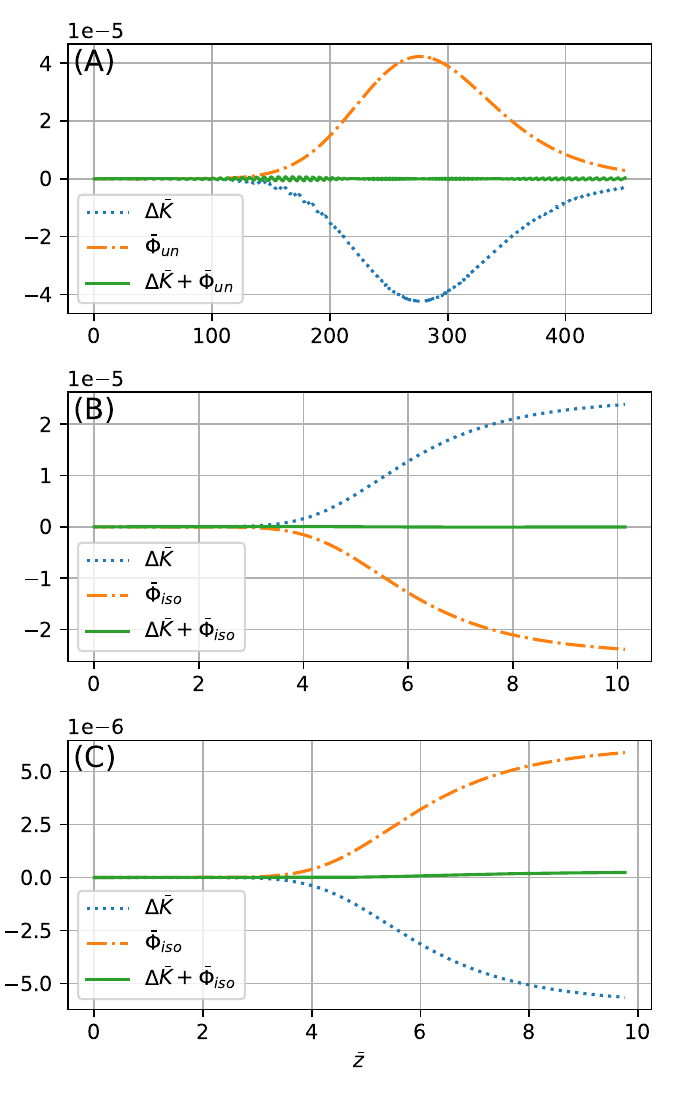}
	\caption{Normalized change in OC kinetic energy $\Delta \bar{K}$, theoretical potential energy $\bar{\Phi}$, and their sum, as a function of distance $\bar{z}$, for cases A, B, and C from Table~\ref{tab:parameters}.
		In (A), $\Phi$ comes from Eq.~(\ref{eq:PhiUnNormalized}), while in (B-C) it comes from Eq.~(\ref{eq:PhiEqNormalized}).
		The agreement to theory is good, and the qualitative aspects of the theory are visible; in particular, the different asymptotic behavior as $\bar{z}\rightarrow \infty$ for the unperturbed vs. isopotential cases, and the switch in the sign of the potential depending on the Doppler-shifted frequency for the isopotential case.}
	\label{fig:PhiVerification}
\end{figure}

This analysis is shown in Fig~\ref{fig:PhiVerification}, for the unperturbed case (A), the low-frequency ($\bar{\omega} < 1$) isopotential case (B), and the high-frequency ($\bar{\omega} > 1$) isopotential case (C).
The agreement to the theory is quite good in each case, and the qualitative features of the solution are clearly visible.
First, $\Phi_\text{un}$ goes to 0 as $\bar{z} \rightarrow \infty$ (Fig.~\ref{fig:PhiVerification}A), while $\Phi_\text{iso}$ goes to a constant (Fig.~\ref{fig:PhiVerification}B-C).
Second, $\Phi_\text{iso}$ is attractive for $\bar{\omega} < 1$ (Fig.~\ref{fig:PhiVerification}B), and repulsive for $\bar{\omega} > 1$ (Fig.~\ref{fig:PhiVerification}C).

\subsection{Limits of Cold Ponderomotive Theory}

However, the assumptions of the theory can break down.
First, when $\bar{k} \langle \bar{v}_\perp \rangle = k \rho$ becomes large, the linear response becomes kinetic \cite{stix1992waves}, violating the cold-plasma assumption in Refs.~\onlinecite{dodinPonderomotiveBarrierMaxwell2004,Dodin2005} and thus changing the form of the ponderomotive potential \cite{Ochs2023PonderomotiveRecoil}.
Second, when $\langle \bar{v}_\perp \rangle / \bar{E}_0 = \langle v_\perp \rangle / v_{E\times B} \gtrsim 1$, a large enough wave amplitude produces a transition to Landau-resonant stochastic diffusion \cite{Karney1978,karney1979stochastic,ochs2021nonresonant}.
The amplitude threshold for this transition to occur can be written as $\bar{\Theta} \gtrsim 1$, where:
\begin{align}
	\bar{\Theta} \equiv 4 \sqrt{\frac{2}{\pi}} \left|\int_0^{\bar{z}} d\bar{z}'\bar{\tilde{B}}(\bar{z})\right| \lp  \frac{ \bar{\omega}}{\langle\bar{v}_\perp \rangle}  \rp^{3/2} \bar{E}_0^{1/2}.
\end{align}
Technically, this threshold only takes this form for $\langle \bar{v}_\perp \rangle / \bar{E}_0 \gg 1$, but it should gives a rough range in which the wave sufficiently dephases from the particle over a single cyclotron period to result in stochastic diffusion.

\begin{figure}
	\centering
	\includegraphics[width=\linewidth]{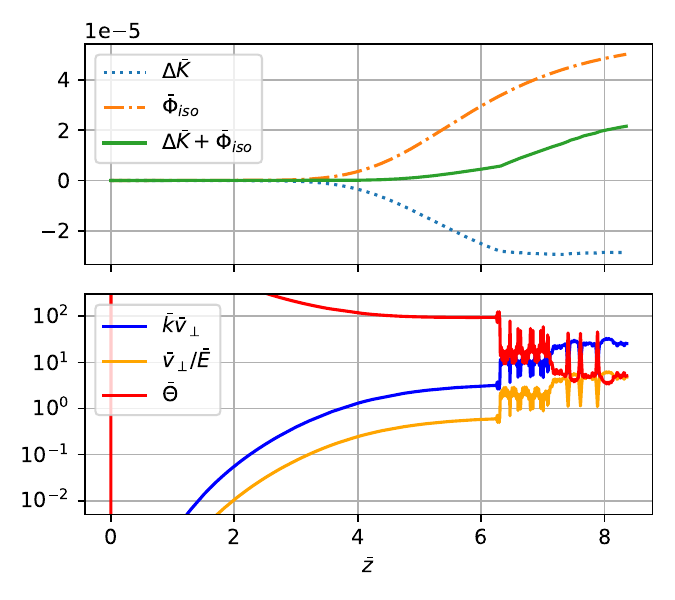}
	\caption{The normalized kinetic and potential energies, and their sum (top), for case D, a particle that violates the small-parameter assumptions of the ponderomotive theory (bottom).
		First, the particle gains enough $\langle \bar{v}_\perp \rangle$ that $k \langle \bar{v}_\perp \rangle > 1$, so that the ponderomotive potential the particle sees is no longer consistent with the cold-plasma expression from Eq.~(\ref{eq:PhiEqNormalized}).
		Second, $\langle \bar{v}_\perp \rangle$ increases enough that $\langle \bar{v}_\perp \rangle / \bar{E}_0 \sim 1$, which (along with the fact that $\Theta > 1$), leads the particle to have a resonant interaction with the wave, resulting in stochastic diffusion in $\langle \bar{v}_\perp \rangle$ and the loss of a ponderomotive potential.}
	\label{fig:PhiEqDeviation}
\end{figure}

An example of a particle which violates these validity criteria is given by case D. 
Its trajectory is shown in the top half of Fig.~\ref{fig:PhiEqDeviation}, and the various dimensionless parameters that determine the behavior in the bottom half.
Initially, the particle satisfies all criteria for the ponderomotive theory to hold, and the agreement is good.
However, as the perpendicular energy increases due to interaction with the wave, $\bar{k} \langle \bar{v}_\perp \rangle$ becomes larger than 1, and the energy begins to diverge.
However, the trajectory is still smooth, implying that the particle is still seeing a ponderomotive potential--just not the \emph{cold-plasma} ponderomotive potential.
However, as $\langle \bar{v}_\perp \rangle$ increases even further, $\langle \bar{v}_\perp \rangle / \bar{E}_0$ becomes $\mathcal{O}(1)$, and (since $\bar{\Theta}$ is already greater than one) there is a sudden onset of strong stochastic diffusion that qualitatively changes the particle behavior, so that it no longer sees the potential.

\section{Discussion and Conclusion}
A very simple slab model demonstrates that in an $\ve{E}\times \ve{B}$-flowing plasma with magnetostatic perturbations, the relationship between the electric potential and the perturbed field lines can completely change the structure of the wave in the plasma rest frame, and thus change the character of the ponderomotive potential experienced by charged particles.
In particular, the plasma-frame waves from isopotential field lines are perpendicularly polarized due to the Lorentz invariance of $\ve{E} \cdot \ve{B}$, with a magnitude that depends on the \emph{integral} of the perturbing field.
In contrast, when the isopotential surfaces are not perturbed by the wave, the plasma frame waves are parallel-polarized, with a magnitude that depends on the local perturbing field amplitude.
We also leveraged several key parameters, familiar from hot-plasma ponderomotive wave theory \cite{stix1992waves,Ochs2023PonderomotiveRecoil} and nonlinear resonance-broadening theory \cite{Karney1978,karney1979stochastic}, to determine the validity limits of the cold-plasma ponderomotive theory.
Single-particle simulations confirmed the theory and its region of validity.

In light of this work, it is clear that, while ponderomotive end plugging represents a significant opportunity, there are several important outstanding problems to address.
First and foremost, of course, is the question of what the proper isopotential structure is,
which depends on a complex wave propagation problem \cite{fetterman2010stationary,fetterman2012wave,Zhmoginov2009WavesAlpha,Zhmoginov2012ApplyingAlphachanneling,Gueroult2023WavePropagation,Langlois2023ContributionFictitious} that in general differs dramatically depending on the specific wave excited in the rotating plasma.
This problem is already complicated even for a weak ponderomotive potential, where the plasma can be treated as approximately uniform.
For the application to a strong ponderomotive potential, as is desirable for end plugging, proper treatment of the behavior of the bulk plasma requires self-consistency between the plasma density and the potential, so that the wave dielectric tensor actually changes within the end plug, possibly changing the nature of the wave.
While this self-consistent problem is complex, the analysis here provides a mapping between the (currently unresolved) wave structure of the perturbation and the resulting ponderomotive potential, with gyroresonances becoming important only when the electric potential structure is perturbed along with the magnetic field structure.

Second, we explored here only a small region in the large space of possible ponderomotive wall geometries, which might have a large impact on their behavior.
As an example, if the slab wall we considered also had a large constant $B_y$ component, a single field line would experience both positive and negative $B_x$, and thus it would not diverge as far from its starting position, which would change the strength of the wall considerably.

Ultimately, what this study exposes is a fundamental subtlety in the theory of ponderomotive forces: namely that the way in which the electric potential relaxes upon perturbed magnetic field lines can have dramatic effects on both the nature and the efficacy of the ponderomotive walls produced by those perturbations. 




%
%
%
%
%
%

\acknowledgments{The authors would like to thank Elijah Kolmes, Tal Rubin, and Jean-Marcel Rax for useful discussions.
This work was supported by ARPA-E Grant DE-AR0001554 and NNSA grant DE-SC0021248.
This work was also supported by the DOE Fusion Energy Sciences Postdoctoral Research Program, administered by the Oak Ridge Institute for Science and Education (ORISE) and managed by Oak Ridge Associated Universities (ORAU) under DOE contract No. DE-SC0014664.}

%


\clearpage
		
\end{document}